\documentclass[prl,twocolumn,aps,superscriptaddress]{revtex4-2}
\usepackage{color}
\usepackage[usenames,dvipsnames,svgnames]{xcolor}
\usepackage{graphicx}
\usepackage{amsmath}
\usepackage{natbib}
\usepackage{graphicx}
\usepackage{mathtools}
\usepackage{float}
\usepackage{mathrsfs}
\usepackage{indentfirst}
\newcommand{\wmk}{~Wm$^{-1}$K$^{-1}$~}
\usepackage{hyperref}
\hypersetup{
pdfnewwindow=true,      % links in new window
colorlinks=true,        % false: boxed links; true: colored links
linkcolor=Blue,         % color of internal links
citecolor=Blue,         % color of links to bibliography
filecolor=Blue,         % color of file links
urlcolor=Blue           % color of external links
}

\begin{document}

%\title{Thermal transport in single wall carbon nanotubes}
\title{Ultrahigh yet convergent thermal conductivity of carbon nanotubes from comprehensive atomistic modeling}

\author{Giuseppe Barbalinardo}
\affiliation{Department of Chemistry, University of California, Davis, Davis, CA 95616, USA}

\author{Zekun Chen}
\affiliation{Department of Chemistry, University of California, Davis, Davis, CA 95616, USA}

\author{Haikuan Dong}
\affiliation{QTF Centre of Excellence, Department of Applied Physics, Aalto University, FI-00076 Aalto, Finland}
\affiliation{College of Physical Science and Technology, Bohai University, Jinzhou, 121013, China}

\author{Zheyong Fan}
\affiliation{QTF Centre of Excellence, Department of Applied Physics, Aalto University, FI-00076 Aalto, Finland}
\affiliation{College of Physical Science and Technology, Bohai University, Jinzhou, 121013, China}

\author{Davide Donadio}
\affiliation{Department of Chemistry, University of California, Davis, Davis, CA 95616, USA}
\email{ddonadio@ucdavis.edu}
\date{\today}

\begin{abstract}
Anomalous heat transport in one-dimensional nanostructures, such as nanotubes and nanowires, is a widely debated problem in condensed matter and statistical physics, with contradicting pieces of evidence from experiments and simulations. 
Using a comprehensive modeling approach, comprised of lattice dynamics and molecular dynamics simulations, we proved that the infinite length limit of the thermal conductivity of a (10,0) single-wall carbon nanotube is finite but this limit is reached only for macroscopic lengths due to thermal phonon mean free path of several millimeters. Our calculations showed that the extremely high thermal conductivity of this system at room temperature is dictated by quantum effects. Modal analysis showed that the divergent nature of thermal conductivity, observed in one-dimensional model systems, is suppressed in carbon nanotubes by anharmonic scattering channels provided by the flexural and optical modes with polarization in the plane orthogonal to the transport direction.
\end{abstract}

%\keywords{Suggested keywords}%Use showkeys class option if keyword
\maketitle

The 1955 computer experiment by Fermi, Pasta, Ulam (FPU) and Tsingou~\footnote{E. Fermi, J. Pasta, S. Ulam, Los Alamos Scientific Laboratory report LA-1940 (1955). Published later in E. Segr\`e, ed., Collected Papers of Enrico Fermi, Vol. 2, U. Chicago Press, Chicago (1965).} initiated the field of nonlinear statistical physics and the age of computer simulations in physics. The FPU model, consisting of a one-dimensional (1D) chain of masses interacting through weakly anharmonic springs, sparked a debate on anomalous heat transport in low-dimensional systems. 
While in standard bulk materials heat diffusion obeys Fourier's law and the thermal conductivity $\kappa$ is an intrinsic property, the thermal conductivity of the FPU and other 1D models diverges with their length $L$ as $\kappa \sim L^\alpha$, with $\alpha> 0$~\cite{Lepri:2003fc,Dhar:2008ij}.
Heat transport in these models is deemed anomalous, as it violates the principles of normal diffusion~\cite{Lepri:1997, Lepri:2005gg, Li:2010kc, Liu:2014jk, Lepri:2016bv, Benenti:2000, PhysRevLett.125.040604}. 

The advent of nanotechnology and the discovery of carbon nanotubes (CNT)~\cite{Iijima:1991wj} provided a suitable platform to verify experimentally the predictions from nonlinear statistical models. 
The early measurements on suspended individual CNTs found thermal conductivity comparable or even higher than diamond~\cite{Hone:1999dm, Kim:2001ei, Pop:2006ih, Balandin:2011gk}.
Systematic measurements of $\kappa$ vs. length in multi-wall and, eventually, single-wall (SW)CNTs supported the hypothesis of anomalous heat transport with diverging $\kappa(L)$ ~\cite{Chang:2008cp, Lee:2017he}.  
The latter experiment showed that for suspended SWCNTs $\kappa$ may reach values as high as 13,000~\wmk\ at room temperature (RT) and would not converge up to millimeter lengths. 
A similar experiment, however, suggests that $\kappa$ converges for pristine suspended SWCNT at lengths of few tens of $\mu$m~\cite{Liu:2017hv}. The technical difficulties of these experiments and possible shortcomings in the models used to interpret measurements~\cite{Li:2017ez} suggest that the fundamental problem of anomalous heat conduction in CNTs and other 1D nanostructures may not be solved experimentally. 

Even for metallic CNTs, the electronic contribution to $\kappa$ is negligible, and the major heat carriers are phonons~\cite{Watanabe:2004}, thus making the problem tractable by either molecular dynamics (MD) or anharmonic lattice dynamics (ALD) and the Boltzmann transport equation (BTE).
Several theoretical and computational studies have tackled the calculation of the thermal conductivity of either finite or infinitely long CNTs, but they do not provide a clear and coherent picture altogether~\cite{Zhang:2020ce}.
Equilibrium MD (EMD) simulations with periodic boundary conditions probe the infinite-length limit, provided that size convergence and phase-space sampling are properly addressed. For CNTs of various chirality modeled with different interatomic potentials, most EMD simulations suggest that $\kappa$ is finite~\cite{Donadio:2007ev, Thomas:2010jq, Pereira:2013dz, Fan:2015ba, Fan:2019gr}.  
Conversely, most nonequilibrium MD (NEMD) simulations 
show no evidence of $\kappa$ converging for length up to 10~$\mu$m~\cite{Zhang:2005bz, Shiomi:2006ku, Saaskilahti:2015ka, Ushnish:2019, Bruns:2020jc}. 
Early ALD-BTE calculations suggested that $\kappa$ would converge only if 4-phonon scattering processes are explicitly considered~\cite{Mingo:2005cz}. Successive works, implementing the self-consistent solution of BTE implied that $\kappa$ may converge only by imposing a cutoff to low-frequency phonon modes~\cite{Lindsay:2009cz, Lindsay:2010ce}.
EMD, NEMD and ALD-BTE have complementary strengths and weaknesses. For example, MD simulations do not approximate the anharmonic terms of the interaction potential, but, as the dynamics is Newtonian, do not include quantum mechanical (QM) effects, which are easily implemented in ALD-BTE, instead.     

In this Letter, we reconcile the results from EMD and NEMD with those from ALD-BTE, to finally provide a coherent and comprehensive picture of heat transport in CNTs.  
By concertedly performing MD and ALD-BTE simulations on a (10,0) SWCNT, we show that the thermal conductivity for this paradigmatic system converges to a large value, comparable to that measured in recent experiments. The convergence length is $\ge 1$~mm, {in accordance with measurement showing length dependence of $\kappa$ for mm-long CNTs}. Our calculations show an overall quantitative agreement among the different simulation methods, provided that they are carefully converged and boundary conditions are implemented correctly. In particular, we find that the Matthiessen rule, customarily used in ALD-BTE calculations, fails spectacularly for CNTs with length from few tens of nm to mm, that is when transport is in the so-called nanoscale regime~\cite{HoogeboomPot:2015gd} between the ballistic and the diffusive limit.

\paragraph{Infinite size limit.} We consider a $(10,0)$ semiconducting SWCNT, for which we model the interatomic interactions using the empirical bond-order Tersoff potential~\cite{Tersoff:1989tr} using the parameters optimized to reproduce the phonon dispersion relations of graphene~\cite{Lindsay:2010kq}.
We first compute the thermal conductivity of an infinitely long SWCNT by both EMD and ALD simulations with periodic boundary conditions. In EMD simulations, $\kappa$ is computed in the Green-Kubo formalism as an infinite time integral of the heat flux auto-correlation function obtained from a sufficiently large ensemble of EMD runs in the microcanonical ensemble~\cite{Green1952,Green1954,Kubo:1957we,ZWANZIG:1965tp}. We performed EMD simulations using the GPUMD package~\cite{GPUMD} for periodic $(10,0)$ SWCNT cells of lengths $52$--$207$ nm at 300~K. While the Green-Kubo integral would not converge for 1D systems with anomalous heat transport~\cite{Lepri:2003fc}, our EMD simulations converge with respect to both correlation time and cell length and provide a value of $\kappa=2408 \pm 83$~\wmk
{(see Supplementary Material, Table ST1, Figures S1, S2)}.

\begin{figure}[t]
\centering
\includegraphics[width=0.9\linewidth]{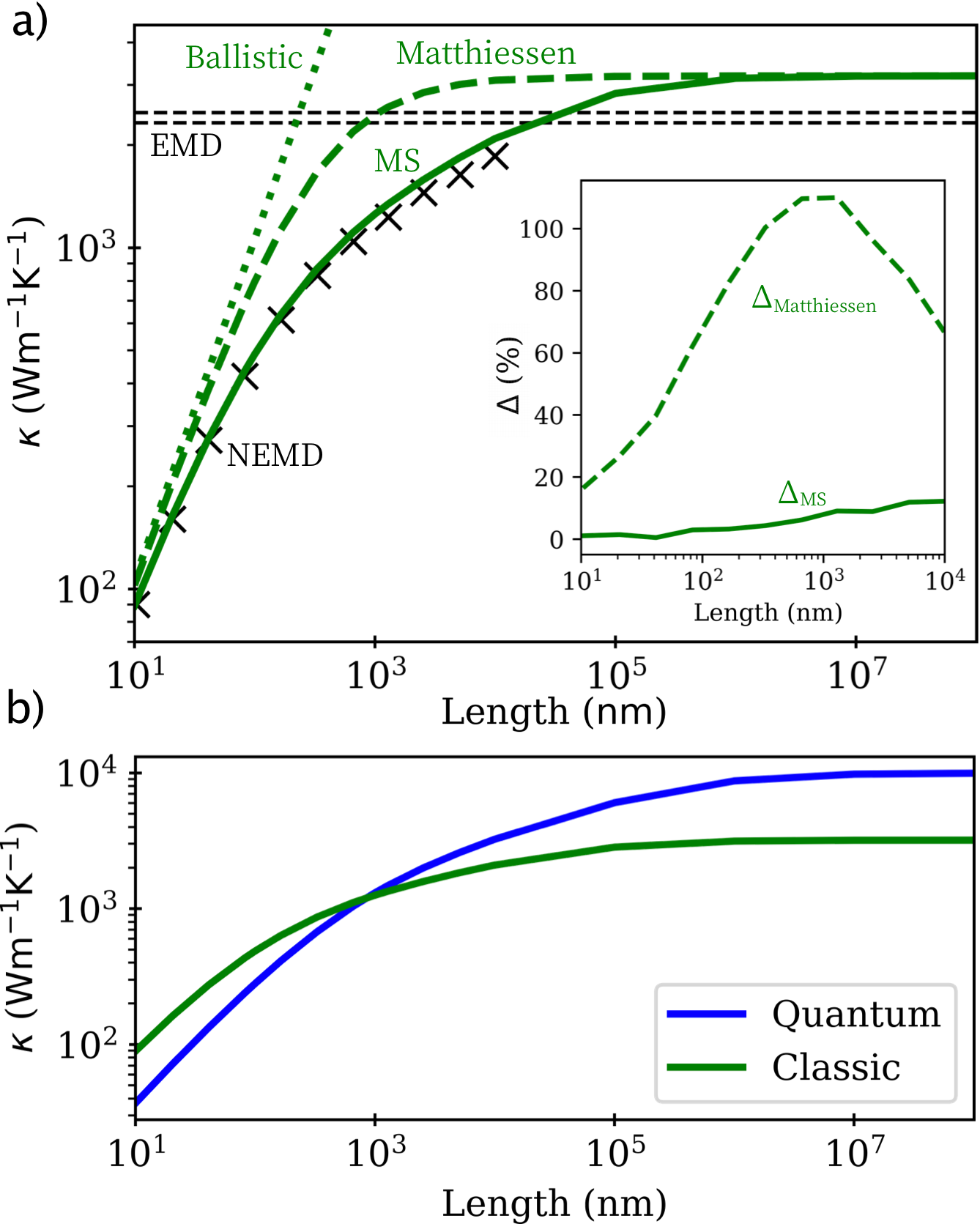}
\caption{
a) BTE and MD classical simulations at 300 K as a function of the CNT length. The BTE finite size corrections are calculated using both the MS method (solid green line) and the Matthiessen rule (dashed green line). The ballistic limit (dotted green line) is plotted as a reference. The NEMD result up to 10 $\mu$m (black x) is shown, as well as the infinite limit EMD, where the double black dashed line represents the error bars. The inset shows the difference in percentage of the finite size BTE, with respect to the NEMD simulations, $\Delta=100\cdot(\kappa_{\rm BTE} - \kappa_{\rm NEMD}) / \kappa_{\rm NEMD}$.
b) Comparison of quantum (blue) and classical (green) simulations using the BTE+MS approach.
}
\label{ald_vs_md_classic}
\end{figure}

Whereas the convergence of $\kappa$ in EMD simulations is supported by plenty of evidence, ALD-BTE calculations have so far struggled to provide an estimate for the thermal conductivity of SWCNT in the infinite length limit. 
ALD-BTE allows one to compute $\kappa$ from the nonequilibrium phonon population  $\delta n_{\mu}=n_{\mu} - \bar n_{\mu}$ obtained by solving the linearized BTE~\cite{Peierls:1929jv}, which for a system extended in one dimension reads:
\begin{equation}\label{eq:bte}
 v_\mu \frac{\partial{n_\mu}}{\partial x}=   {v}_\mu  \frac{\partial \bar n_\mu}{\partial T}\nabla T = - \sum_{\mu'}\Gamma_{\mu\mu'} \delta n_{\mu'},
\end{equation}
where $\Gamma_{\mu\mu'}$ is the scattering tensor that accounts for all the phonon scattering processes~\cite{srivastava1990physics}. Here we consider only anharmonic scattering approximated to the first order of anharmonicity, i.e. considering only three-phonon scattering processes. The resulting thermal conductivity is   
\begin{equation}\label{eq:kappaBTE}
\kappa = \frac{1}{N_{q} V} \sum_{\mu} c_{\mu} v_{\mu} \lambda^{\textrm{pp}}_{\mu},
\end{equation}
where $N_q$ is the number of $q$ points used to integrate the first Brillouin zone, $c_\mu$ and $v_{\mu}$ are the heat capacity and group velocity of mode $\mu$~\footnote{The index $\mu$ contracts both the polarization and q-point index.}. 
Due to prominent hydrodynamic effects~\cite{Cepellotti:2015ke,Lee:2015ex}, the phonon mean free path (MFP) $\lambda^{\textrm{pp}}_{\mu}$ cannot be approximated by the relaxation time approximation, and needs to be computed by inverting the scattering tensor as $\lambda^{\textrm{pp}}_{\mu}=\sum_{\mu'}(\Gamma^{-1})_{\mu\mu'} v_{\mu'}$.
Matrix inversion is often circumvented by a self-consistent approach~\cite{Omini:1995jl,Broido:2005kf} that was applied in the previous ALD-BTE calculations for SWCNT~\cite{Lindsay:2009cz, Lindsay:2010ce}. 
However, the self-consistent series converges only if the off-diagonal elements of the scattering tensor are smaller than the diagonal ones in modulus {$|\Gamma_{\mu\mu'}/\Gamma_{\mu\mu}|<1$  with $\mu\neq \mu'$.~\cite{Cepellotti:2016bk}}
This is not the case for the low-frequency modes of the (10,0) SWCNT (Figure S5), hence we directly invert the scattering tensor~\cite{Barbalinardo:2020bu}, with well-converged $q$-point sampling (Figure S4)
\footnote{$\kappa$ is computed using the $\kappa$ALDo software and a system of 5 replicas of the 40 atoms CNT fundamental unit cell. The second and third order interatomic force constants are calculated by finite displacement using the USER-PHONON package in LAMMPS~\cite{Plimpton:1995wl}. The Brillouin zone is sampled using a 200 $q$-points reciprocal space mesh, which has been tested to be converged. The second and third-order derivatives of the interatomic force constants are obtained by finite difference displacement, as explained in \cite{Barbalinardo:2020bu}.}.
%and using the full range of the short-range interatomic potential to evaluate the third derivatives. 
%Diffferent ways of solving BTE~\cite{Omini:1995jl,Broido:2005kf, Cepellotti:2015ke, Barbalinardo:2020bu}. 
 ALD-BTE presents the advantage that it can be carried out both with the correct QM phonon statistics and in the classical limit, which is taken by solving Eq.~(\ref{eq:bte}) with $\hbar\rightarrow 0$. 
Both classical and QM calculations converge but to very different values of $\kappa$. The classical limit at 300~K gives $\kappa_{cl}=3190$~\wmk which compares well to the value computed by EMD, especially considering that in ALD-BTE anharmonic effects are truncated, leading to overestimated $\kappa$ even at low temperature~\cite{Turney:2009bb,Feng:2016ib,Gu2019PRB}
The QM calculation gives $\kappa_{QM}=9,960$~\wmk in the same ballpark as the experiments on mm-long SWCNTs~\cite{Lee:2017he}. 
The difference between $\kappa_{QM}$ and $\kappa_{cl}$ is among the largest for any material at RT: it stems from both the high Debye temperature $\Theta_D$ and the low dimensionality of CNTs. 
Such striking difference indicates that classical simulations for carbon-based nanomaterials can be used to find qualitative trends, but cannot provide quantitative predictions. \textcolor{black}{Quantum effects become less important at high temperature or for materials with lower $\Theta_D$, for which MD  provides reliable results.}.  

\paragraph{Finite size.}  NEMD allows one to study thermal transport in an analogous way to the experiments. Two local thermostats apply a temperature difference $\Delta T$ at opposite sides of a system with finite length $L$ to generate a nonequilibrium heat flux $J$~\cite{MullerPlathe:1997ub,Jund:1999tb,Shiomi2014NONEQUILIRIUMMD}. 
Evaluating ${J}$ at a stationary state and probing the temperature difference $\Delta T$ between the two reservoirs, {with temperature controlled through Langevin dynamics,} provides the thermal conductance per unit area $G=J/\Delta T$ and the effective length-dependent thermal conductivity $\kappa(L)=GL$~\cite{Li:2019fo,Hu:2020eq}. \textcolor{black}
Here we performed NEMD simulations to compute $\kappa(L)$ for the $(10,0)$ SWCNT with $L$ from  10 nm to 10 $\mu$m using GPUMD \cite{GPUMD} {exploiting a simulation protocol carefully evaluated for carbon nanostructures~\cite{Li:2019fo} (see SM, Fig. S3, Tab. S1, including Refs.~\cite{Nose:1984wa,Martyna:1992gy})}.
The NEMD results displayed in Fig.~\ref{ald_vs_md_classic}a show that $\kappa(L)$ apparently diverges with length, at least up to 10~$\mu$m, which is at the edge of the length accessible to NEMD simulations. Also these results are in agreement with former works, but they do not contradict the former EMD predictions, as $\kappa(L)$ remains below the infinite limit $\kappa(\infty)$ computed by EMD. 

To compute the length-dependent thermal conductivity  $\kappa(L)$ of the SWCNT at QM level, we use the ALD-BTE method. The customary way of introducing finite length in ALD-BTE is to correct the anharmonic phonon MFPs with a boundary scattering term through Matthiessen's rule~\cite{Ziman1960}:
$1/\lambda^{L}_{\mu} = 1/\lambda^{\textrm{diff}}_{\mu} + 1/{L}$,
which correctly describes both the ballistic ($L\rightarrow 0$) and the diffusive ($L\rightarrow \infty$) limits.
Figure~\ref{ald_vs_md_classic}a shows a large discrepancy between ALD-BTE results obtained using the Matthiessen rule and the reference NEMD results: for any $L>10$~nm the ALD-BTE calculations systematically overestimate $\kappa(L)$. 
The reason for the failure of the Matthiessen rule is that it does not account correctly for the boundary conditions for Eq.~(\ref{eq:bte}). Specifically, in Eq.~(\ref{eq:bte}) the first equality is justified under the assumption of local thermal equilibrium $\frac{\partial n_\mu}{\partial x}= \frac{\partial n_\mu} {\partial T} \nabla T= \frac{\partial \bar n_\mu}{\partial T} \nabla T$, which is not the case in systems with finite length, \textcolor{black}{as suggested by the nonlinear temperature profiles of our NEMD simulations (Figure S3)}. 
Using the McKelvey-Shockley (MS) flux method, Maassen and Lundstr\"om proposed a solution of the BTE for finite systems with consistent boundary conditions~\cite{Maassen:2015ih}. This method gives up the assumption of local thermal equilibrium and treats separately forward and backward fluxes along the transport direction. 
The finite-length thermal conductivity resulting from this approach for a 1D systems is:
\begin{equation}\label{eq:mckelvey_shockley}
\kappa(L)=\frac{1}{N_{q} V} \sum_{\mu\mu^{\prime}} c_{\mu} v_{\mu}\left(\Gamma_{\mu\mu'} +  \frac{2|v_{\mu^{\prime}}|}{L} \delta_{\mu\mu'} \right)^{-1}v_{\mu^\prime}
\end{equation}
in which the boundary scattering term  $L/(2|v_{\mu^{\prime}}|)$ is added to the diagonal elements of $\Gamma_{\mu\mu'}$. 
$\kappa(L)$ computed using the MS method in the classical limit is in excellent agreement with the reference NEMD calculations (Figure~\ref{ald_vs_md_classic}a).  
The agreement between carefully implemented NEMD simulations and well-converged ALD-BTE calculations with the correct boundary conditions gives solid ground to the finite-size MS approach~\cite{Hu:2020eq} and resolves a long-standing puzzle about  nanoscale heat transport.

\begin{figure}[tb]
\centering
\includegraphics[width=0.9\linewidth]{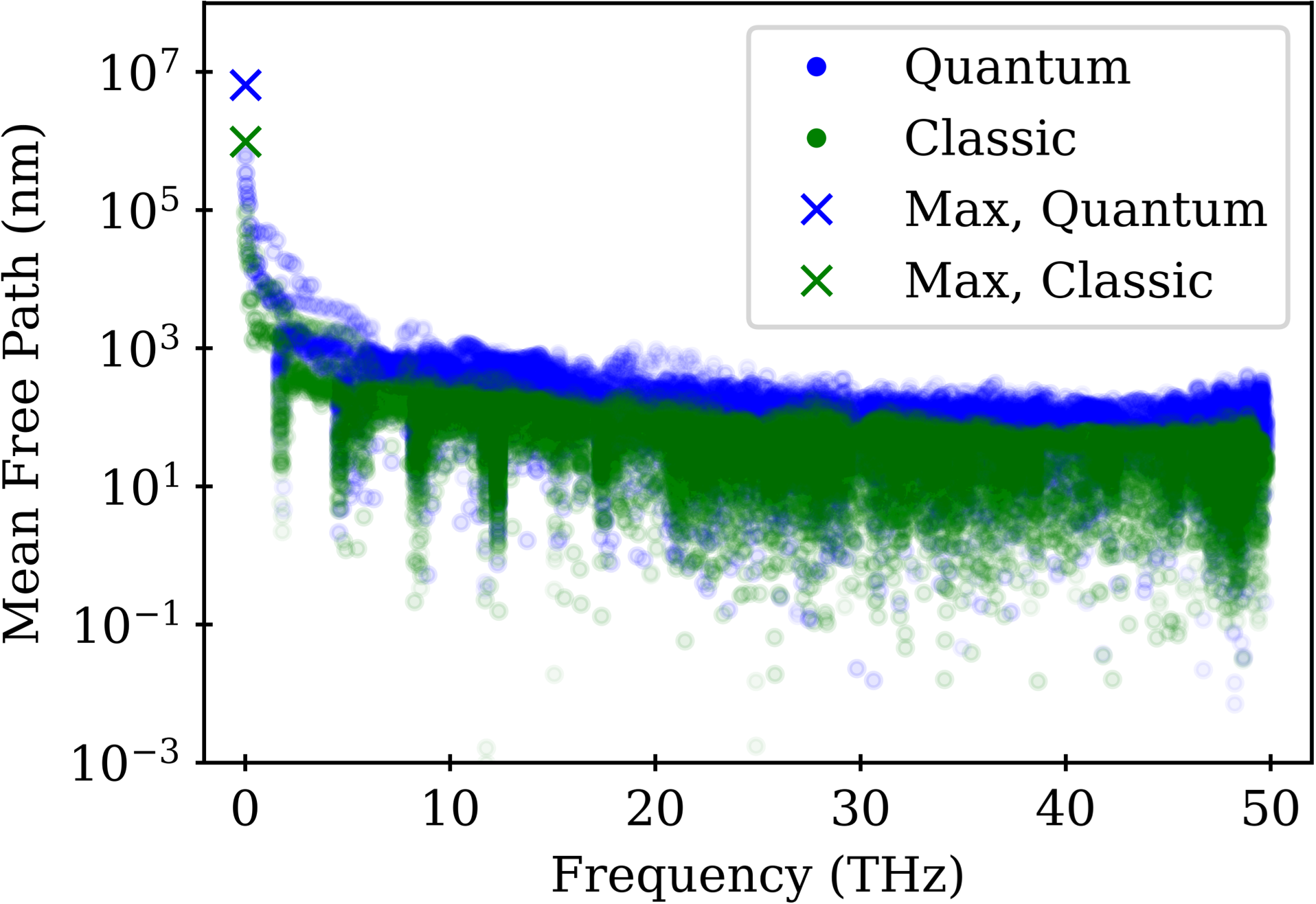}
\caption{Phonon mean free paths as a function of the frequency, computed using QM (blue) and classical (green) statistics. The $(\times)$ symbols indicate the longest mean free paths, which are 6~mm (QM) and 0.9~mm (classical).}
\label{mfp_vs_freq}
\end{figure}

We can now exploit ALD-BTE to calculate $\kappa(L)$ with quantum statistics to obtain predictions directly comparable to experiments for any length (Figure~\ref{ald_vs_md_classic}b). 
Quantum effects influence both the modal heat capacity and the MFPs (see Equations~\ref{eq:kappaBTE} and ~\ref{eq:mckelvey_shockley}). Due to equipartition in classical statistical mechanics $c_\mu$(classical) is always larger than  $c_\mu$(QM). Conversely, QM MFPs are longer than the classical ones (Figure~\ref{mfp_vs_freq}), because the number of phonon scattering processes is lower in QM as high frequency modes are less populated. For this reason, the RT $\kappa_{QM}$ is much larger than $\kappa_{cl}$. At finite lengths, however, we observe a crossover at $L\sim$1~$\mu$m, below which $\kappa(L)_{cl}>\kappa(L)_{QM}$ as the effect on modal heat capacity dominates over that on phonon MFPs, which are limited by the length of the system. 
Furthermore, Figure~\ref{ald_vs_md_classic}b shows that the (10,0) SWCNT must be longer than the maximum thermal phonon MFP, i.e. few millimeters, for $\kappa(L)$ to converge to the infinite size limit. 
These results reconcile equilibrium and nonequilibrium simulations, ALD-BTE calculations and experimental measurements, all painting the same coherent picture of thermal transport in SWCNTs: 
the infinite-length limit of $\kappa$ is finite, but such limit is reached for mm-long SWCNTs, which are difficult to probe experimentally. 

%%%%%%% MODAL ANALYSIS

\begin{figure}[t]
\centering
\includegraphics[width=1.\linewidth]{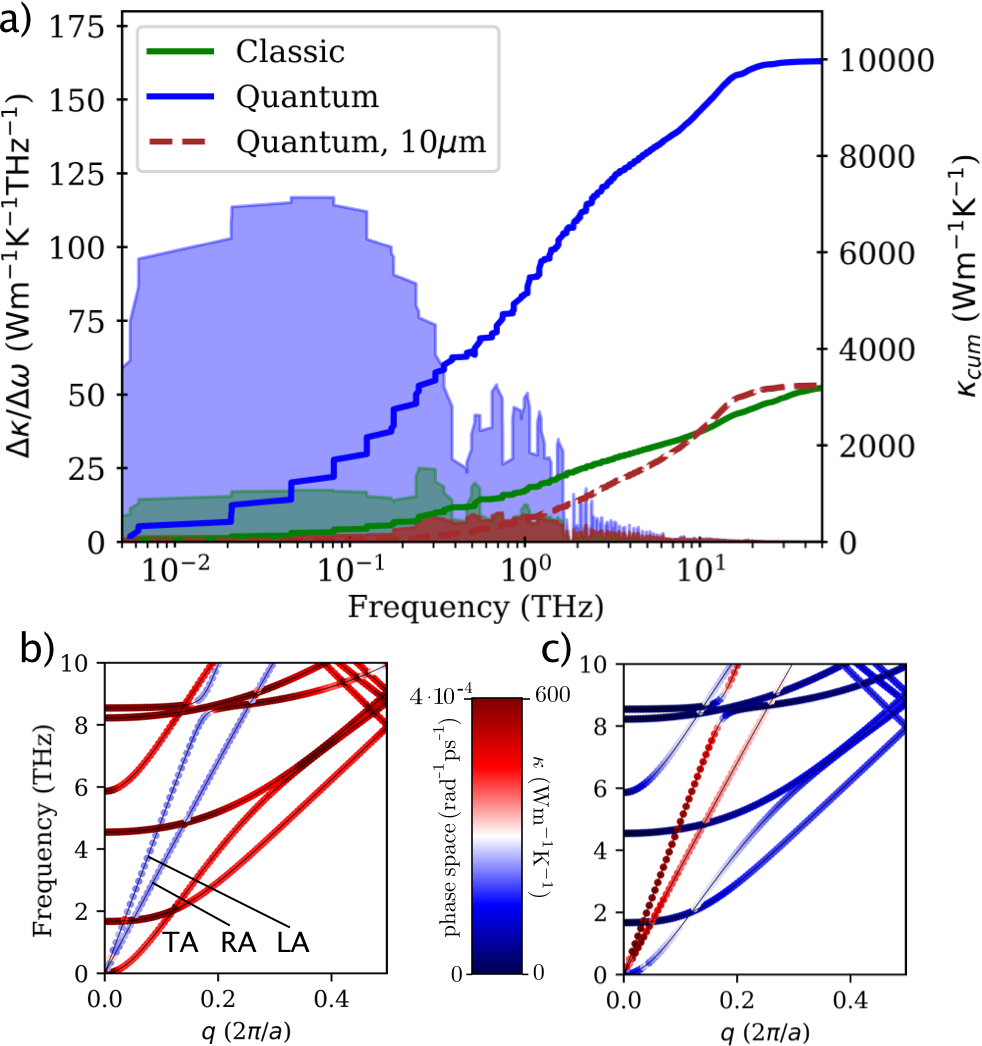}
\caption{\label{fig:conductivity_per_mode_300}
a) Frequency resolved conductivity at 300~K, using both the quantum and the classical statistics. The solid lines show the diffusive cumulative conductivity (reference axis on the right), while the shaded area shows the differential one (reference axis on the left), for quantum (blue), and classical (green). The brown shaded area and dashed lines show the quantum case for a $10~\mu m$-long CNT. Panels (b) and (c) show the dispersion relations below 10 THz, colored according to their contribution to the scattering phase space (b), and thermal conductivity as in Eq.~\ref{eq:kappaBTE} (c). {The labels indicate the transverse acoustic (TA),  torsional acoustic (RA), and longitudinal acoustic (LA) modes.}}
\end{figure}
\paragraph{Modal analysis.} Hereafter, we analyze the origin of the large difference between $\kappa_{cl}$ and $\kappa_{QM}$, and why heat transport in SWCNTs is not anomalous as opposed to 1D models. %, i.e. $\kappa$ converges to finite value at infinite length. 
Figure~\ref{fig:conductivity_per_mode_300}a displays the comparison between the classical and QM frequency-resolved thermal conductivity $\kappa(\omega)$ for the infinite-length SWCNT. The shaded areas represent the differential contribution $\Delta\kappa/\Delta\omega$.
The contribution to $\kappa_{QM}$ from frequencies lower than 1 THz amounts to about 50\% of the total, and that below 10~THz to 90\%.
In the classical calculation the contribution from the low-frequency modes is substantially reduced, as they are involved in a larger number of anharmonic scattering processes. The difference stems from the 2-to-1 phonon scattering processes that involve one high-frequency and one low-frequency mode combining into another high-frequency phonon.  %(or at high temperature)  
Classical statistics overestimates these processes due to the excess population of high-frequency modes, which are depleted according to the correct QM statistics.  
Figure~\ref{fig:conductivity_per_mode_300}b and c show the dispersion relations near the $\Gamma$ point with the color of the dots indicating the 3-phonon scattering phase space~\cite{Li:2015jq}, i.e. the number of scattering processes available to each phonon (b) and the mode-resolved contribution to the total thermal conductivity (c). 
The longitudinal and torsional acoustic (LA and RA) modes are involved in relatively few scattering processes and provide the largest contribution to $\kappa$. The two degenerate transverse acoustic modes (TA) also substantially contribute to $\kappa$, but they also provide a large number of scattering channels. Conversely, higher order flexural modes have a large scattering phase space without contributing to heat transport mostly because of their small group velocity.   
Figure~\ref{fig:conductivity_per_mode_300}a also shows the finite length effect on the relative modal contribution to $\kappa$. As the long MFP of low-frequency modes is truncated by the system boundaries, the main contribution to $\kappa$ shifts toward higher-frequencies modes with shorter intrinsic $\lambda_{pp}$. 

%%%%% 1D SYSTEM

\begin{figure}[b]
\centering
\includegraphics[width=0.9\linewidth]{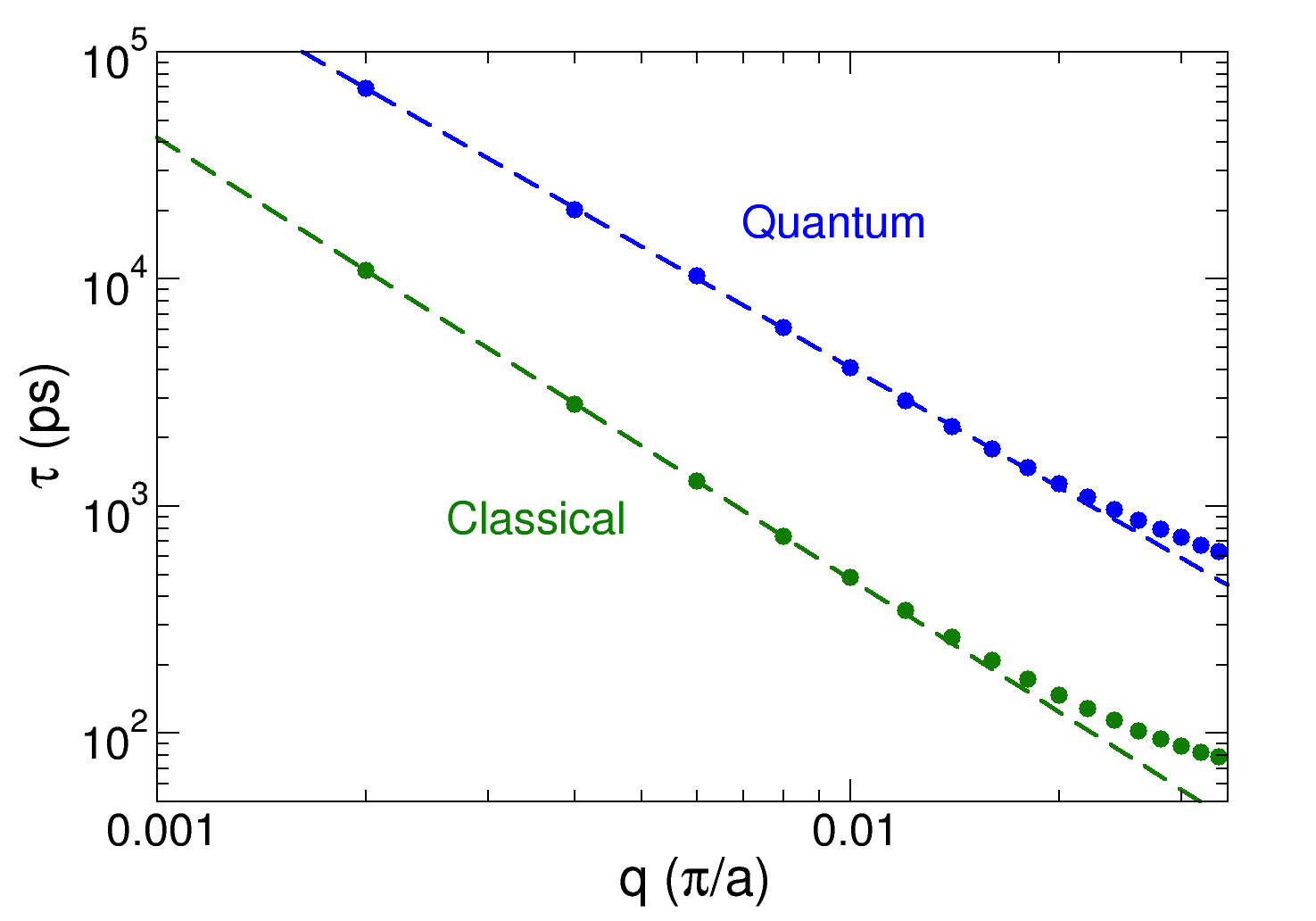}
\caption{\label{fig:1dCNT} Lifetime $\tau$ of the longitudinal acoustic mode as a function of the wavevector $q$ for the SWCNT in which atoms can move only along the symmetry axis. $\tau$ is computed by BTE with either quantum or classical phonons statistics. The low-$q$ limit is fitted to $\tau \propto q^{-\eta}$ (dashed lines), with $\eta=1.76$ (quantum) and 1.94 (classical). 
}
\end{figure}
\paragraph{Effect of dimensionality.}
Finally, we investigate the difference between the SWCNT, a 3D system extended in one dimension with converging $\kappa(L)$, and a system with the same structure in which the atomic coordinates perpendicular to the CNT axis are constrained, so that the dynamics is 1D. 
The phonon dispersion relations of this model, compared to those of the 3D SWCNT are displayed in Figure~S7. 
ALD-BTE calculations suggest that the lifetime of the longitudinal acoustic mode diverges for $q\longrightarrow 0$ as $\tau\propto q^{-\eta}$, with $\eta=1.76$ (quantum) and 1.96 (classical), thus supporting the divergence of $\kappa(L)\propto L^\alpha$ with  $\alpha=0.43$ (quantum) and 0.48 (classical). Trends and exponents are compatible with both classical and quantum lattice dynamics calculations of anomalous transport in the FPU model~\cite{Pereverzev:2003jx, Santhosh:2007kz},  showing that the suppression of the dynamics in the plane orthogonal to the transport direction restores anomalous thermal transport with $\kappa\propto L^\alpha$ as in 1D models.

\paragraph{Conclusions.}
We showed that different atomistic approaches, MD and ALD-BTE, independently provide the evidence that the thermal conductivity of SWCNTs has a finite bulk limit, in contrast with anomalous transport observed in 1D models \textcolor{black}{in 3D space~\cite{Barreto2019}.} 
% suspended polymer threads~\cite{Henry:2008kg,Crnjar:2018fg} and aligned atomic chains~\cite{Yang:2021ec}.}
This result does not conflict with experimental observations of increasing $\kappa(L)$ for mm-long nanotubes, as the computed convergence length at RT is beyond 1~mm. 
To capture the correct magnitude of $\kappa$ it is essential to treat phonons quantum mechanically, meaning that classical MD simulations can provide qualitative trends but not accurate predictions below $\Theta_D$. 
For the (10,0) SWCNT considered here, the infinite length limit of $\kappa$ at the QM level is about three larger than in the classical limit. 
Modal analysis shows that both the two fundamental and the optical higher order flexural modes provide the main source for anharmonic scattering channels that make the thermal conductivity of SWCNTs finite.
Our work reconciles formerly divergent numerical approaches to compute thermal transport in CNTs, and highlight the fundamental differences between 1D models that showcase anomalous heat transport, and nanostructures with finite thermal conductivity in the bulk limit, thus providing a substantial contribution to solve a critical problem across condensed matter and statistical physics. We expect this fundamental result to hold not only for CNTs, but also for nanowires and other nanomaterials that exhibit strong dependence of $\kappa$ on length at the macroscopic scale~\cite{Hsiao:2013gz}. \textcolor{black}{Divergence may still occur for suspended polymer threads~\cite{Henry:2008kg,Crnjar:2018fg} and aligned atomic chains~\cite{Yang:2021ec}.}

\begin{acknowledgments}
We are grateful to Stefano Lepri, Shunda Chen and Giuliano Benenti for stimulating discussions. G.B. acknowledges support by the MolSSI Investment Software Fellowships (NSF grant No. OAC-1547580-479590).
H.D. and Z.F. acknowledge support by the National Natural Science Foundation of China under Grant No. 11974059.
{
The data that support this study are openly available in Zenodo at 10.5281/zenodo.4698465~\cite{zenodo2021}.}
%Data are available  at \textcolor{blue}{https://github.com/nanotheorygroup/kaldo}.
\end{acknowledgments}

\bibliography{ultimate}

\end{document}